\documentclass[a4paper,12pt]{article}

\usepackage[a4paper,top=2.0cm,bottom=2.0cm,left=2.0cm,right=2.0cm,headheight=14.5pt]{geometry}
\usepackage[vlined,titlenumbered,linesnumbered,algo2e,ruled]{algorithm2e} 
\usepackage{float}
\usepackage{amsmath,amssymb}
\usepackage{amsfonts}
\usepackage{mathtools}
\usepackage{fancyhdr}
\usepackage{setspace}
\usepackage{enumitem}
\usepackage[comma,numbers]{natbib}
\usepackage[mathlines]{lineno}
\usepackage{colortbl}
\usepackage{tikz}
\usetikzlibrary{arrows,matrix,backgrounds,calc,snakes,patterns,chains}

\title{\bf A note on searching sorted unbalanced three-dimensional arrays}
\author{{\bf M\'arcia R. Cappelle}\\[-1pt]
             { \bf Les Foulds}\\[-1pt]
             { \bf Humberto J. Longo}\\
        {\footnotesize Instituto de Inform\'atica}\\[-4pt]
        {\footnotesize Universidade Federal de Goi\'as}\\[-4pt]
        {\footnotesize Alameda Palmeiras, Quadra D, Campus Samambaia}\\[-4pt]
        {\footnotesize CEP 74001970, Goi\^ania--GO, Brazil}\\
        {\footnotesize {\tt \{marcia,lesfoulds,longo\}@inf.ufg.br}}
       }
\date{}

\makeatletter
\let\inserttitle\@title
\let\insertauthor\@author
\makeatother
\begin{document}
\thispagestyle{empty}
\maketitle

\begin{abstract}
We examine the problem of searching sequentially for a desired real value (a key) within a sorted unbalanced three-dimensional finite real array. This classic problem can be viewed as determining the correct dimensional threshold function from a finite class of such functions within the array, based on sequential queries that take the form of point samples. This note addresses the challenge of constructing algorithms that require the minimum number of queries necessary in the worst case, to search for a given key in arrays that have three dimensions with sizes that are not necessarily equal.\\[10pt]
\end{abstract}

\section{Introduction}
Searching is one of the most basic, frequent and important operations performed in tasks involving computation. A search is used to decide whether or not a given value (termed a key) is present among a given collection of values. When the search must be performed sequentially (as is assumed throughout in this paper) and repeatedly, obviously it should be carried out as efficiently as possible. The complexity of such a search is completely dependent on how the collection is organized. Clearly, when there is no information available about collection organization, every value must be examined. However, when the collection is already sorted, the search can be conducted in a far more efficient manner. It is well-known that binary  \citep{knuth3-1973} and saddleback \citep{dijkstra-1976} search algorithms have the best possible worst-case performance for the sequential search of one-dimensional and of two-dimensional square arrays, respectively. The purpose of this note is to describe a sequential search algorithm for the case of three-dimensional arrays that have sizes that are not necessarily equal. This important search problem occurs in many computation-related fields, including computational biology and image processing \citep{baase-gelder-1999,cosnard-duprat-ferreira-1989,gries-1981,sarnath-he-1992}. 

Consider a finite real $3$-dimensional array $A = \{a(i_1,i_2,i_3)\mid i_1 = 1,\dots,n_1;i_2 = 1,\dots,n_2;i_3 = 1,\dots,n_3\}$, that is sorted, i.e., satisfies $a(i_1,i_2,i_3) \leqslant a(j_1,j_2,j_3)$ if $i_1 \leqslant j_1$, $i_2 \leqslant j_2$ and $i_3 \leqslant j_3$ for given, independent, $n_1,n_2,n_3 \in \mathbb{N}$, (so, entries of $A$ are nondecreasing along its dimensions). The array $A$ can be embedded in the $3$-dimensional space with entry $a(1,1,1)$, (one of the minimum entries of $A$) located at the origin, termed the Southwest (SW) corner and with entry $a(n_1,n_2,n_3)$, (one of the maximum entries of $A$) located at the Northeast (NE) corner.

The purpose of this note is to examine the search of $A$ sequentially in order to establish, as efficiently as possible, whether or not a given key $x \in \mathbb{R}$ is a member of $A$. If $x$ is found the search is terminated, as it is assumed throughout that there is no necessity to identify further possible instances of the key. It is assumed here that the comparisons between $x$ and the members of $A$ must be conducted sequentially. A comparison of $x$ and any element $a(i_1,i_2,i_3) \in A$ returns exactly one of the three following results:
\begin{align}
x &< a(i_1,i_2,i_3), \label{cmp-a-1}\\
 \intertext{in which case the entries $a(j_1,j_2,j_3)$ for $j_1 \geqslant i_1$, $j_2 \geqslant i_2$ and $j_3 \geqslant i_3$ can be discarded, or}
x &= a(i_1,i_2,i_3), \label{cmp-a-2}\\
 \intertext{in which case the key has been located and the search is terminated, or}
x &> a(i_1,i_2,i_3), \label{cmp-a-3}\\
 \intertext{in which case the entries $a(j_1,j_2,j_3)$ for $j_1 \leqslant i_1$, $j_2 \leqslant i_2$, and $j_3 \leqslant i_3$ can be discarded.}\nonumber
\end{align}

Suppose $A$ is an $n_1\times n_2\times n_3$ matrix with $n_1 \geqslant n_2 \geqslant n_3 \geqslant 1$. Due to its unbalanced dimensions sizes, the array $A$ is termed \emph{tower}. It is assumed from now on that $n_3 \geqslant 4$. (Otherwise, a simple algorithm based on applying binary search sequentially to each of the $m$ rows of the smallest size of $A$ is worst-case optimal.)

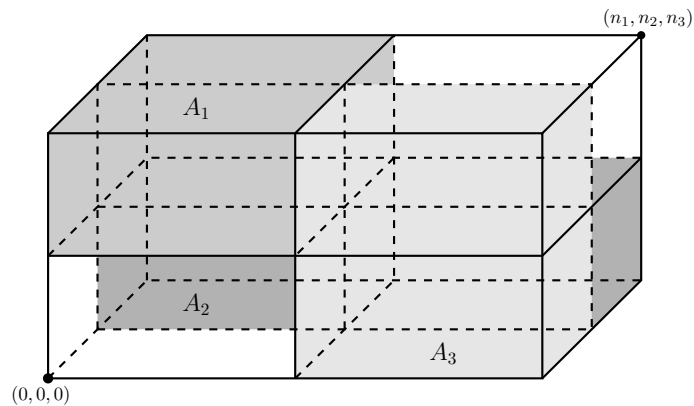
\begin{figure}[h!]
  \centering
  \begin{tikzpicture}[
    ->,>=stealth',thick,
    every node/.style={font=\large\bfseries},
    transform shape,
    scale=0.65,
   ]
   \coordinate (A1) at ( 0, 0);
   \coordinate (A2) at ( 0, 5);
   \coordinate (A3) at (10, 5);
   \coordinate (A4) at (10, 0);
   \coordinate (B1) at ( 1, 1);
   \coordinate (B2) at ( 1, 6);
   \coordinate (B3) at (11, 6);
   \coordinate (B4) at (11, 1);
   \coordinate (C1) at ( 2, 2);
   \coordinate (C2) at ( 2, 7);
   \coordinate (C3) at (12, 7);
   \coordinate (C4) at (12, 2);
   \coordinate (D1) at ( 5, 0);
   \coordinate (D2) at ( 5, 5);
   \coordinate (D3) at ( 7, 7);
   \coordinate (D4) at ( 7, 2);
   \coordinate (E1) at ( 0,2.5);
   \coordinate (E2) at ( 2,4.5);
   \coordinate (E3) at (12,4.5);
   \coordinate (E4) at (10,2.5);
   \coordinate (F1) at ( 1,3.5);
   \coordinate (F2) at ( 7,4.5);
   \coordinate (F3) at (11,3.5);
   \coordinate (F4) at ( 5,2.5);
   \coordinate (G1) at ( 6,1.0);
   \coordinate (G2) at ( 6,6.0);

   \draw[-,dashed] (B1) -- (B2) -- (B3) -- (B4) -- (B1);
   \draw[-,dashed] (A1) -- (C1) -- (C4);
   \draw[-,dashed] (C1) -- (C2);
   \draw[-,dashed] (D1) -- (D4) -- (D3);
   \draw[-,dashed] (E1) -- (E2) -- (E3);
   \draw[-,dashed] (F1) -- (F3);
   \draw[-,dashed] (F2) -- (F4);
   \draw[-,dashed] (G1) -- (G2);

   \draw[-] (A1) -- (A2) -- (A3) -- (A4) -- (A1);
   \draw[-] (A2) -- (C2) -- (C3) -- (C4) -- (A4);
   \draw[-] (A3) -- (C3);
   \draw[-] (D1) -- (D2) -- (D3);
   \draw[-] (E1) -- (E4) -- (E3);

   \draw[fill=black] (A1) circle (0.2em) node[below,xshift=-4pt] {\small$(0,0,0)$};
   \draw[fill=black] (C3) circle (0.15em) node[above,xshift=4pt] {\small$(n_1,n_2,n_3)$};
   \node at ($(A2) + (3.0,0.5)$) {$A_1$};
   \node at ($(B1) + (2.0,0.5)$) {$A_2$};
   \node at ($(D1) + (3.0,0.5)$) {$A_3$};
   \begin{scope}[on background layer]
    \fill[black!20!white] (E1) -- (A2) -- (C2) -- (D3) -- (F2) -- (F4) -- (E1);
    \fill[black!10!white] (D1) -- (D2) -- (G2) -- (B3) -- (B4) -- (A4) -- (D1);
    \fill[black!30!white] (B1) -- (1.0,2.5) -- (F4) -- (5.0,1.0) -- (B1);
    \fill[black!30!white] (B4) -- (11.0,4.5) -- (E3) -- (C4) -- (B4);
    \end{scope}
  \end{tikzpicture}
  \caption{Partition of $A$ into  $A_1$, $A_2$,  and $A_3$.}
 \label{fig:mahl}
\end{figure}

We propose the following search algorithm that iteratively searches particular subtowers of smaller sizes after discarding entries of $A$ that cannot be $x$. With regard to the SW-to-NE diagonal of any cube of any tower, consider the maximum entry that is less than or equal to the key $x$. We term this entry the pivot of the cube. The proposed algorithm identifies a pivot by applying binary search to the SW-to-NE diagonal of the middle cube of maximum possible volume of the current subtower. Once found, the pivot is used to discard entries of $A$, if possible, using \eqref{cmp-a-1} and \eqref{cmp-a-3}. The remainder of the tower is then divided into three smaller subtowers $A_1$, $A_2$ and $A_3$ (see Figure \ref{fig:mahl}) and the process is repeated on these subtowers until either $x$ is found or all of the remaining subtowers are one- or two-dimensional subarrays. Binary search is used to explore any remaining vectors and Bird's method \citep{bird-2006} is used for any two-dimensional subarrays.

The algorithm proceeds by calling the subroutine MAHL-$e(A(1,\dots,n_1;1,\dots,n_2;1,\dots,n_3),\allowbreak x)$. The variables $\ell_i$ and $r_i$, $i =1,2,3$, in this subroutine, denote the limits of the search area of the original tower at each recursive step; $p^1=(p^1_1,p^1_2,p^1_3)$ and $p^2=(p^2_1,p^2_2,p^2_3)$ denote the limits of the current SW-to-NE diagonal; and $c=(c_1,c_2,c_3)$ is the pivot of the current the SW-to-NE diagonal of each cube searched. The logical operator ``$\vee$'' in line \ref{diagonal3dlinha15} indicates that if $x$ is found in any of the recursive calls of the algorithm, then the algorithm is teminated.

Lines \ref{diagonal3dlinha3} through \ref{diagonal3dlinha13} of the algorithm represent a binary search on the SW-to-NE diagonal, the recursive steps are given in line \ref{diagonal3dlinha15}, and in line \ref{diagonal3dlinha17} the subroutine \textit{1$d$-2$d$-Method()} is called. To simplify the description of the algorithm, we assume that this subroutine performs the searches needed for vectors or two dimensional arrays.

\begin{center}
\begin{algorithm2e}[H]\small
 \caption{MAHL-$e(A, x)$}
 \DontPrintSemicolon
 \BlankLine
 \KwIn{Array $A(\ell_1,\dots,r_1;\ell_2,\dots,r_2;\ell_3,\dots,r_3)$ and key $x\in \mathbb{R}$.}
 \KwOut{True if $x\in A$ or False otherwise.}
 \BlankLine
 $found \leftarrow \mathrm{False}$;\;
 \eIf{$(\ell_1 < r_1)\> \mathrm{\bf and}\> (\ell_2 < r_2)\> \mathrm{\bf and}\> \ell_3 < r_3)$}
{
\tcp*[l]{\footnotesize Binary search on the SW-to-NE diagonal.}
    $p^1 \leftarrow (\ell_1,\ell_2,\ell_3)$;\;\nllabel{diagonal3dlinha3}
    $p^2 \leftarrow (r_1,r_2,r_3)$;\;
 	\While{$(p^1_1 \leqslant p^2_1)\> \mathrm{\bf and}\> (p^1_2 \leqslant p^2_2)\> \mathrm{\bf and}\> (p^1_3 \leqslant p^2_3)\> \mathrm{\bf and}\> (found=\mathrm{False})$}
 	{
		$c \leftarrow \left(\lfloor (p^1_1+p^2_1)/2 \rfloor, \lfloor (p^1_2+p^2_2)/2 \rfloor,\lfloor (p^1_3+p^2_3)/2 \rfloor\right)$;\;
		\eIf{$(x>A(c_1,c_2,c_3))$}
		{
			$p^1_i \leftarrow c_i+1,\> i=1,2,3$;\;
		}
		{
			\eIf{$(x<A[c_1,c_2,c_3])$}
			{ 
				$p^2_i \leftarrow c_i-1,\> i=1,2,3$;\;
			}
			{
				$found \leftarrow \mathrm{True}$;\;\nllabel{diagonal3dlinha13}
			}
		}
 	}
    \BlankLine
	\If{$(found=\mathrm{False})$}
 	{\nllabel{diagonal3dlinha14}
        \tcp*[l]{Recursive call of the algorithm.}
		$\begin{array}{rcl}
		found& \leftarrow & \text{MAHL-}e(A(\ell_1,\dots,r_1;p^1_2,\dots,r_2;\ell_3,\dots,p^2_3),x)\> \vee\>\\
		      &                 & \text{MAHL-}e(A(\ell_1,\dots,p^2_1;\ell_2,\dots,r_2;p^1_3,\dots,r_3),x)\> \vee\>\\
		      &                 & \text{MAHL-}e(A(p^1_1,\dots,r_1;\ell_2,\dots,p^2_2;\ell_3,\dots,r_3),x);\;
		  \end{array}$
		\nllabel{diagonal3dlinha15}
 	}
    \BlankLine
}
{
  \tcp*[l]{$(\ell_1 = r_1)\> \mathrm{\bf or}\> (\ell_2 = r_2)\> \mathrm{\bf or}\> \ell_3 = r_3)$}
  \tcp*[l]{\footnotesize Search on a subarray of one or two dimensions.}
 $found \leftarrow$ 1$d$-2$d$-Method$(A(\ell_1,\dots,r_1;\ell_2,\dots,r_2;\ell_3,\dots,r_3),x)$;\;		\nllabel{diagonal3dlinha17}
}
\Return{$found$.}
\end{algorithm2e}
\end{center}

Let $\tau(n_1, n_2, n_3)$ be the number of comparisons necessary to determine whether or not $x$ occurs in an $n_1\times n_2\times n_3$ tower $A$. The number of elements in the SW-to-NE diagonal is equal to the lowest dimension of the array. Therefore, $\tau(n_1, n_2, n_3)$ is the sum of $\lg(n_3+1)$ (induced by the search of the SW-to-NE diagonal of the central cube) plus the comparisons required in the remaining subtowers. The recurrence relations for the proposed algorithm MAHL-$e$ are:
\begin{equation}\label{eq-mahl-rec}
  \tau(n_1,n_2,n_3) =
   \begin{cases}
   1, & \text{if } n_1=n_2=n_3 = 1;\\
   \lg(n_1+1), & \text{if }n_1>1 \text{ and } n_2=n_3=1;\\
   \lg(n_2+1), & \text{if }n_2>1 \text{ and } n_1=n_3=1;\\
   \lg(n_3+1), & \text{if }n_3>1 \text{ and } n_1=n_2=1;\\
   \mathcal{O}(n_1\lg(n_2/n_1+1)), & \text{if } n_1,n_2 > 1 \text{ and } n_3 =1;\\
   \mathcal{O}(n_1\lg(n_3/n_1+1)), & \text{if } n_1,n_3 > 1 \text{ and } n_2 =1;\\
   \mathcal{O}(n_2\lg(n_3/n_2+1)), & \text{if } n_2,n_3 > 1 \text{ and } n_1 =1;\\
    \mathcal{O}(\lg(\min\{n_1,n_2,n_3\}+1)\\
    \qquad +\> \tau(n_1, \lceil n_2/2\rceil, \lfloor n_3/2\rfloor) & \\
    \qquad +\> \tau(\lfloor n_1/2\rfloor, n_2, \lceil n_3/2\rceil)&\\
    \qquad +\> \tau(\lceil n_1/2\rceil, \lfloor n_2/2\rfloor,n_3),& \text{otherwise.}
   \end{cases}
\end{equation}

\section{Conclusions and future work}
We examined the problem of searching sequentially for a desired real value (a key) within a sorted unbalanced three-dimensional finite real array. This classic problem was viewed as determining the correct dimensional threshold function from a finite class of such functions within the array, based on sequential queries that take the form of point samples. This note addressed the challenge of constructing algorithms that require the minimum number of queries necessary in the worst case, to search for a given key in arrays that have three dimensions with sizes that are not necessarily equal.

As future work we intend to present worst case optimal algorithms for unbalanced case including the solution of the recurrence (\ref{eq-mahl-rec}). 


\bibliographystyle{apalike}
\bibliography{multisearch}

\end{document}